\def\kms{\,{\rm km}\,{\rm s}^{-1}}
\def\hmpc{\,{h^{-1} {\rm Mpc}}}
\def\mpch{\,{h {\rm Mpc}^{-1}}}
\def \der{\,{\rm d}}

\documentclass[usenatbib]{mn2e}
\usepackage{epsfig}

\usepackage{amssymb}
\usepackage{amsmath}
\usepackage{widetext}

\begin{document}

\title[Maximum likelihood of cosmic bulk flow]{%
An estimation of local bulk flow with the maximum-likelihood
method}

\author[Y.-~Z.~Ma and J.~Pan]{Yin-Zhe Ma$^{1,2,3\dagger}$ and Jun Pan$^{3,4 \ddagger}$\\
$^1$Department of Physics and Astronomy, University of British Columbia, Vancouver, BC V6T 1Z1, Canada.\\
$^2$Canadian Institute for Theoretical Astrophysics, Toronto, Canada.\\
$^3$2 National Astronomical Observatories, Chinese Academy of
Sciences, 20A Datun Rd., Chaoyang District, Beijing 100012, P.~R.~China.\\
$^4$Purple Mountain Observatory, 2 West Beijing Rd., Nanjing
210008, P.~R.~China.\\
emails: $^{\dagger}$mayinzhe@phas.ubc.ca;\,
$^{\ddagger}$jpan@bao.ac.cn}

\maketitle

\begin{abstract}

A maximum-likelihood method, tested as an unbiased estimator from
numerical simulations, is used to estimate cosmic bulk flow from
peculiar velocity surveys. The likelihood function is applied to
four observational catalogues (ENEAR, SFI++, A1SN and SC)
constructed from galaxy peculiar velocity surveys and Type-Ia
supernovae data at low redshift ($z \leq 0.03$). We find that the
Spiral Field {\it I}-band catalogue constrains the bulk flow to be
$V=290 \pm 30 \kms$ towards $l=281^{\circ} \pm 7^{\circ}$,
$b=8^{\circ +6^{\circ}}_{-5^{\circ}}$ on effective scales of $ 58
\hmpc$, which is the tightest constraints achievable at the
present time. By comparing the amplitudes of our estimated bulk
flows with theoretical prediction, we find excellent agreement
between the two. In addition, directions of estimated bulk flows
are also consistent with measurements in other studies.

\vskip 0.1 truein


\vskip 0.3 truein
\end{abstract}


\section{Introduction}
\label{intro} Cosmic bulk flow is the coherent motion of galaxies
and galaxy clusters towards a particular direction. Since
magnitude and direction of bulk flow are determined by the
underlying density field at large scales, it serves as a direct
probe of the large-scale  structure of the Universe. There have
been a lot of recent studies focusing on estimating bulk flow from
a variety of observational probes, such as galaxy peculiar
velocity survey
\citep{Sarkar07,Springob07,Watkins09,Feldman10,Nusser11}, Type Ia
supernovae data \citep{Sandage10,Dai11,Turnbull12} and galaxy
clusters with observations of the cosmic microwave background
(CMB) radiation \citep{Kashlinsky10,Kashlinsky11}. However,
amplitudes and directions of bulk flows at different depths in our
local Universe obtained from different measurements do not reach
good convergence. Some works have argued that the amplitude of the
bulk flows they found is too high compared to the standard
$\Lambda$ cold dark matter ($\Lambda$CDM) predictions
\citep{Watkins09,Feldman10,Kashlinsky10,Macaulay11,Macaulay12},
which has stimulated a lot of interest in looking for possible
explanation in new physics
\citep{Afshordi09,Mersini-Houghton09,Ma11}.


However, any analysis which claims to strongly rule out the simple
inflationary $\Lambda$CDM model should be subject to careful
examination, since a confirmed large-scale flow would have
profound impact on our understanding of the large-scale structure
of the Universe. \citet{Watkins09} and \citet{Feldman10} adopted
the minimal variance weighting method to estimate bulk flows from
their combined galaxy catalogues, declaring discovery of an excess
power of flow $V=407 \pm 81 \kms$ towards $l=287^{\circ} \pm
9^{\circ}$, $b=8^{\circ} \pm 6^{\circ}$ on a Gaussian window of
$50\hmpc$ (corresponds to a top-hat window function of $\sim 100
\hmpc$). But, by correcting Malmquist bias, selecting high-quality
samples, and combining different data sets with the Bayesian
hyper-parameter method, \cite{Ma13} found that there is no real
excess power of flow on $50 \hmpc$ ($V \sim 310 \kms$,
$l=280^\circ\pm 8^\circ$, $b=5^{\circ}.1 \pm 6^\circ$), and the
estimated amplitude of density fluctuation
$\sigma_8=0.65^{+0.47}_{-0.35}(\pm 1 \sigma)$ is consistent with
{\it Wilkinson Microwave Anisotropy Probe} (\textit{WMAP}) 7-yr
results \citep{Komatsu11}. In \cite{Ma13}, the minimal variance
method is extended to include bulk flows in shells at different
distances ($20$--$100\hmpc$) and a likelihood function is
formulated to combine all of these reconstructed shell velocities,
the multishell likelihood method yields constraints on
cosmological parameters of $\sigma_{8}=1.01^{+0.26}_{-0.20}$ and
$\Omega_{\rm{m}}=0.31^{+0.28}_{-0.14}$ (based on the Spiral Field
{\it I}-band catalogue, in abbreviation SFI++), which are
consistent with {\it WMAP} 7-yr results very well
\citep{Komatsu11}.  The recent estimation of bulk flow based on
the `First Amendment' compilation of 245 Type Ia supernovae found
that bulk flow in the nearby Universe (a Gaussian window of $ 58
\hmpc$) is of $249\pm 76 \kms$ in the direction $l=319^\circ \pm
18^\circ$, $b=7^\circ\pm 14^\circ$ \citep{Turnbull12}, which is in
good agreement with the expectation for the $\Lambda$CDM model
($V\sim 250 \kms$).

Although the detailed analysis with the minimal variance and
multishell likelihood methods in \cite{Ma13} is in itself already
a strong support to disperse the suspicion of a very large local
bulk flow, it is still worthwhile to apply a different method to
the same set of catalogues to check the robustness and reliability
of the reconstructed bulk flow, and test the consistency between
different methods. In this paper, we will use a different bulk
flow reconstruction method, aka the maximum-likelihood method to
calculate the bulk flows of several peculiar velocity catalogues.
Furthermore, we will compare the reconstructed flows with the
theoretical prediction for the $\Lambda$CDM cosmology model and
investigate the tendency of the cosmic flow as a function of
sample depths with currently available peculiar velocity
catalogues.

This paper is organized as follows. Theoretical prediction of bulk
flow on various depths $R$ and the maximum-likelihood method are
presented in section~\ref{sec:bulkflow}. Introduction to the
peculiar velocity samples and Malmquist bias correction is in
section~\ref{sec:dataset}, together with specification on
calculating effective depth of a sample from the geometry of the
peculiar velocity survey. Section~\ref{sec:results} shows the
results of the constraints on the cosmic bulk flows by applying
the likelihood function to the velocity catalogues, and the
comparison against theoretical predictions. Our conclusion is in
the last section.

Throughout the paper, we assume a spatially flat cosmology with
\textit{WMAP} 7-yr best-fitting parameter values
\citep{Komatsu11}, i.e. fractional matter density $\Omega_{\rm
m}=0.2735$, fractional baryon density $\Omega_{\rm b}=0.0455$,
Hubble constant $h=0.704$, power-law index of scalar power
spectrum $n_{\rm s}=0.967$ and amplitude of fluctuation
$\sigma_{8}=0.811$.

\section{Bulk flow model}
\label{sec:bulkflow}

\subsection{Theoretical prediction}
\label{sec:theory} In the linear theory of structure formation,
the velocity field $\mathbf{v}(\mathbf{ r},t)$ is related to the
underlying density field by \cite{Peebles93}
\begin{equation}
\mathbf{v}(\mathbf{r,}t)=\frac{f(t)H(t)}{4 \pi }
 \int_{\rm{all\text{ } space}} \der^{3} \mathbf{r^{\prime }}\delta_{\rm m}(\mathbf{r^{\prime },}t)
 \frac{\mathbf{r}- \mathbf{r^{\prime }}}
 {|\mathbf{r}-\mathbf{r^{\prime }}|^{3}}\ ,
\label{pecu_def}
\end{equation}%
where $\delta_{\rm m}(\mathbf{r})=(\rho (\mathbf{r})-\overline{\rho })/%
\overline{\rho }$ is the density contrast at position
$\mathbf{r}$, $f(t)=d\log D(t)/d \log a(t)\simeq \Omega^{4/7}_{\rm
m}+(1+\Omega_{\rm m}/2)\Omega_{\Lambda}/70$ is the logrithmic
derivative of the linear growth rate \citep{LahavEtal1991,
Dodelson03} and $H(t)$ is the Hubble parameter. Since the bulk
flow we investigate is the streaming motion of very nearby
objects, and the samples are within distance of $150 \hmpc$ of our
local volume, we take the cosmic time `$t$' in
Eq.~(\ref{pecu_def}) to be our present time $t_{0}$, thus the
Hubble parameter becomes Hubble constant $H_{0}$ at $a=1$. The
bulk flow $\mathbf{V}$ is the coherent motion of observed galaxies
or galaxy clusters. Mathematically, $V$ is the velocity field
filtered by the window function defined by the geometry of the
observational sample, and is actually determined by mass
distribution outside the sample space
\citep{Juszkiewicz90,NusserDavis1994, Li12}. The
root-mean-square(hereafter rms) of the bulk motion
($V_{\rm{rms}}$) on scale of $R$ is the velocity power spectrum
filtered by the observational window function \citep{Coles02}
\begin{eqnarray}
V_{\rm{rms}}^{2}(R)  &=&  \langle |
\mathbf{V}(\mathbf{r},t_{0},R)|^{2} \rangle \nonumber \\
 &=& \frac{1}{(2\pi )^{3}}\int
P_{\rm vv}(k)W^{2}(kR)\der^{3}k\ , \label{sigmav}
\end{eqnarray}
where the $W(kR)$ is the Fourier transform of the real space
selection function with size $R$. In linear regime, the velocity
power spectrum at present epoch $a=1$ is \citep{Sarkar07}
\begin{equation}
P_{\rm vv}=\frac{(H_{0} f(t_{0}))^2}{k^{2}} P(k)\ , \label{Pvv2}
\end{equation}
where the $P(k)$ is the linear matter power spectrum which in our
calculation is generated by the software package {\scshape camb}
\citep{Lewis00}. Substituting Eq.~(\ref{Pvv2}) into
Eq.~(\ref{sigmav}) and adopting the simple top-hat window function
$W(x)=3(\sin x-x \cos x)/x^{3}=3 j_{1}(x)/x$ (where $j_{1}(x)$ is
the first spherical Bessel function), the rms of bulk velocity in
a spherical region $R$ becomes (see also \cite{Ma12b} for
derivation)
\begin{eqnarray}
V_{\rm{rms}}^{2}(R) &=& \frac{(H f_{0})^{2}}{(2\pi )^{3}} \int
W^{2}(kR)\frac{P(k)}{k^{2}} \der^{3}k \nonumber \\
&=& \frac{(3H f_{0})^{2}}{2\pi^{2}}\int P(k)
 \left( \frac{j_{1}(kR)}{kR}\right)^{2}\der k\ .
\label{sigmav2}
\end{eqnarray}
Equation (\ref{sigmav2}) is the filtered velocity power spectrum
in real space, which retains large-scale modes of perturbations.
The bulk velocity rms of wider window is smaller than the one for
narrower size window, because more modes are smeared out. Typical
rms of bulk velocity $V_{\rm{rms}}$ in $\Lambda$CDM model from
top-hat window at $20 \hmpc$ is $\sim 350 \kms$, while at $60
\hmpc$ is $\sim 240 \kms$.

Now, given the filtered velocity rms on scale of $R$
(Eq.~(\ref{sigmav2})), what is the probability distribution of the
bulk flow magnitude on this scale? To address this question, we
start from the 3D probability function of the bulk flow velocities
in Cartesian coordinate. The Cartesian components of bulk flow
should be Gaussian distributed, with zero means and variances of
$V_{{\rm rms},x},V_{{\rm rms},y},V_{{\rm rms},z}$ respectively,
assuming null correlation between the three components, the
probability distribution function of bulk flow $\mathbf{V}$ is
\begin{eqnarray}
p(\mathbf{V}) &=& p(V_{x},V_{y},V_{z}) \nonumber \\
& \propto& \exp \left[- \frac{1}{2} \left( \sum_{i=x,y,z}
\left(\frac{V_{i}}{V_{\textrm{rms},i}} \right)^{2}
 \right) \right]. \label{pvxvyvz}
\end{eqnarray}
In an isotropic and homogeneous universe, the velocity field
possesses the property of $V_{{\rm rms},x}^{2}=V_{{\rm
rms},y}^{2}=V_{{\rm rms},z}^{2}=V_{\rm{rms}}^{2}/3$, therefore the
probability of bulk flow with magnitude $V$ becomes
\citep{Bahcall94,Coles02}
\begin{eqnarray}
& p(V)\der V &= \int \left[ p(\mathbf{V})d\Omega _{V}\right] V^{2}\der V \nonumber  \\
 &= & \sqrt{\frac{54}{\pi }}\left( \frac{V}{V_{\rm{rms}}}\right) ^{2}\exp \left[
-\frac{3}{2}\left( \frac{V}{V_{\rm{rms}}}\right) ^{2}\right] \left( \frac{\der V}{%
V_{\rm{rms}}}\right)  ,
\label{pveq}
\end{eqnarray}%
where the final line of equation is properly normalized. So the
amplitude of the bulk flow actually follows the Maxwell-Boltzmann
distribution, which is skewed and has long tail on the large
velocity branch. The peak of the distribution is $V_{\rm
p}=\sqrt{2/3}V_{\rm{rms}}$, which is obtained by taking $\der
p(V)/\der V=0$. One can also calculate the asymmetric variance of
velocities on different depths \citep{Li12}.

In Fig.~\ref{fig:Constraints}, we plot the peak and $\pm 1 \sigma$
variance of the bulk velocity magnitude as a function of scale $R$
in solid line and dashed lines respectively. One can clearly see
that the bulk motion amplitude decreases with increasing $R$. This
is because for the top-hat window function $W(x)\simeq1$ if
$x\ll1$, but $W(x)\simeq 0$ if $x>1$, the upper limit in
Eq.~(\ref{sigmav}) is $ R^{-1}$. Therefore, a large volume (large
$R$) would result in a relatively small value of bulk flow rms. In
addition, the smaller the scale is, the larger the variance of the
bulk flow is. This is the effect of the sample variance, because
if the velocity field is filtered on a smaller scale, larger
variance of this filtered velocity will be. If one averages the
peculiar velocity over the whole Universe ($R \rightarrow \infty
$), the average velocity should be fairly close to zero, if the
primordial perturbations are adiabatic Gaussian as assumed in the
concordance $\Lambda$CDM model \footnote{\cite{Turner91} and
\cite{Ma11} discussed the bulk flows with isocurvature initial
conditions, in which case the overall non-zero average velocity,
aka tilted universe, is possible within such scenario.}.

We need to mention that, the model of Equations~(\ref{Pvv2}) and
(\ref{sigmav2}) for peculiar velocity and subsequent bulk flow is
made under the `single-particle' assumption, i.e. the galaxies do
not strongly correlate with each other and therefore
Maxwellian-Boltzmann distribution can be used to describe its
behaviour. Since for our peculiar velocity catalogues, the data
are quite sparse and not very correlated on small scales, our
assumption is a good approximation \footnote{The average distance
between two objects for the four catalogues is close to $60 \hmpc$
, in which case the correlation energy only takes around $10.3$
per cent of the total kinetic energy.}. On the other hand, in the
regime where gravitational clustering of the galaxies and their
collisions cannot be negligible, one needs to look in to the scale
dependence of the small-scale modes and then consider the
correlation between small and large scales (i.e. the gravitational
quasi-equilibrium distribution method
\citep{Raychaudhury96,Ahmad02,Leong04,Sivakoff05,Saslaw10,Saslaw00}).
Since we are most interested in large-scale bulk flows of which
the small-scale velocity dispersion is smoothed out, we will not
get involved into details of gravitational clustering properties
in this paper.

\subsection{The maximum-likelihood method}
\label{sec:likelihood}

\begin{figure*}
\centerline{\includegraphics[bb=0 0 458
308,width=3.0in]{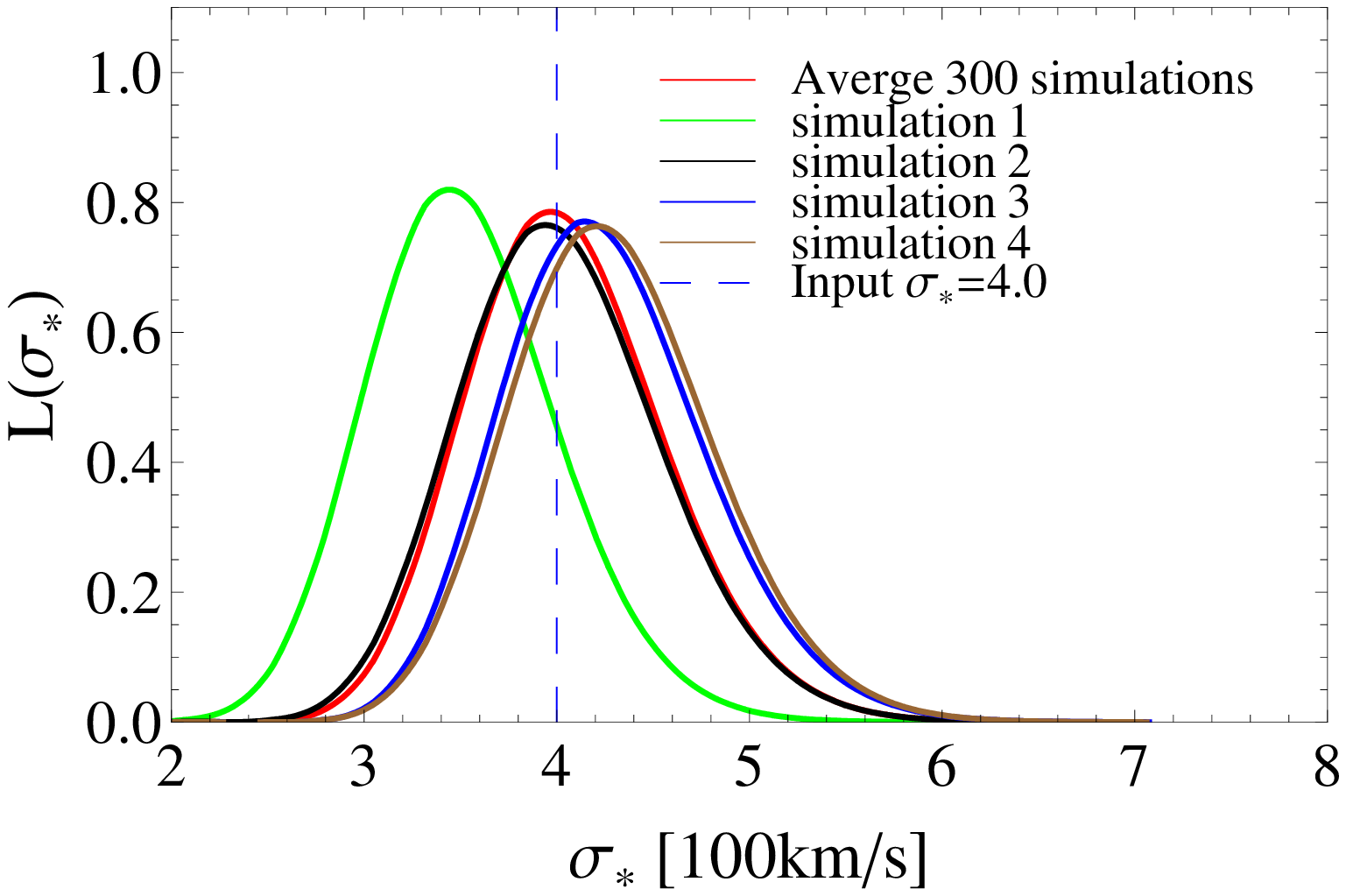}
\includegraphics[bb=0 0 450 300,width=3.0in]{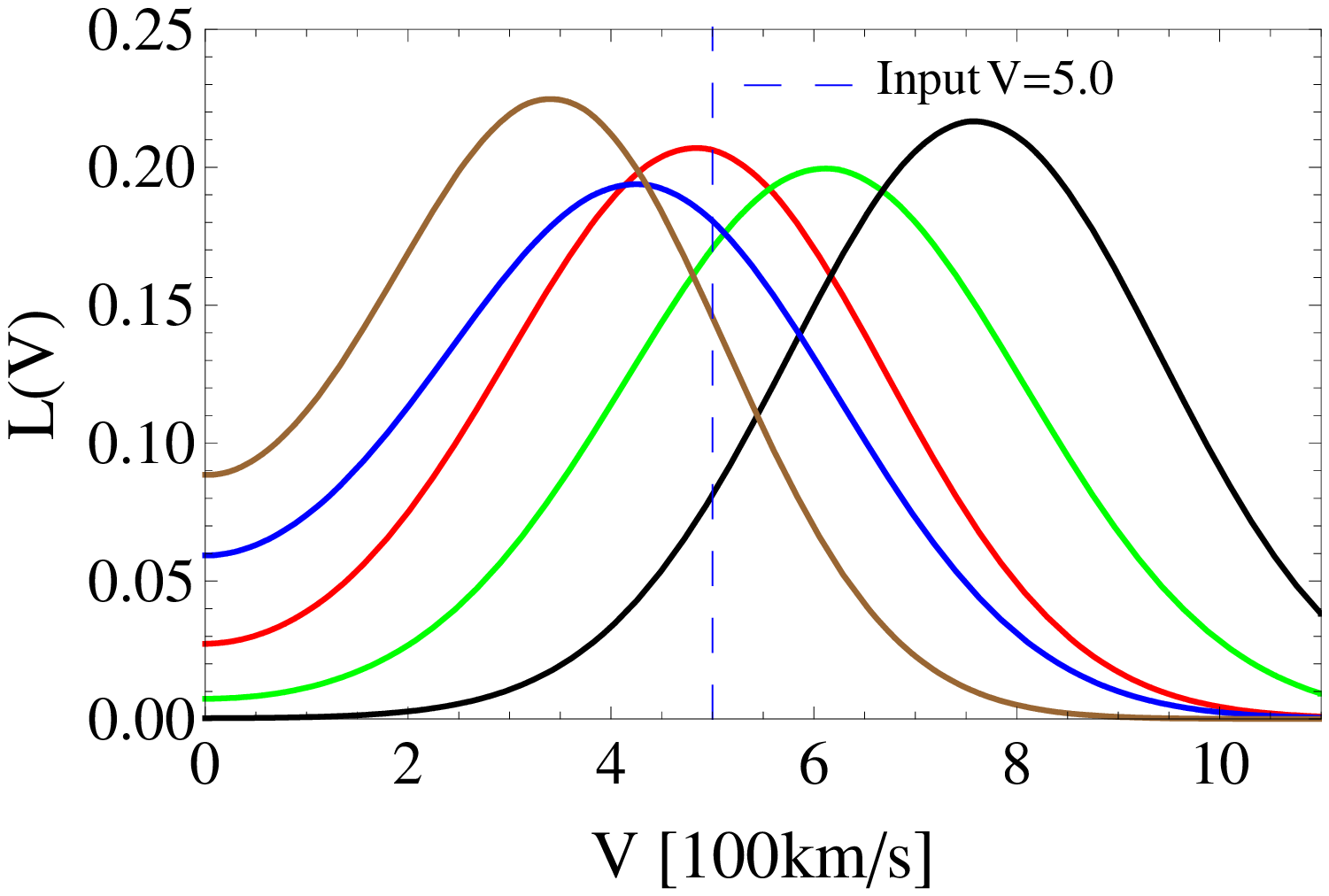}}
\centerline{\includegraphics[bb=0 0 405
275,width=3.0in]{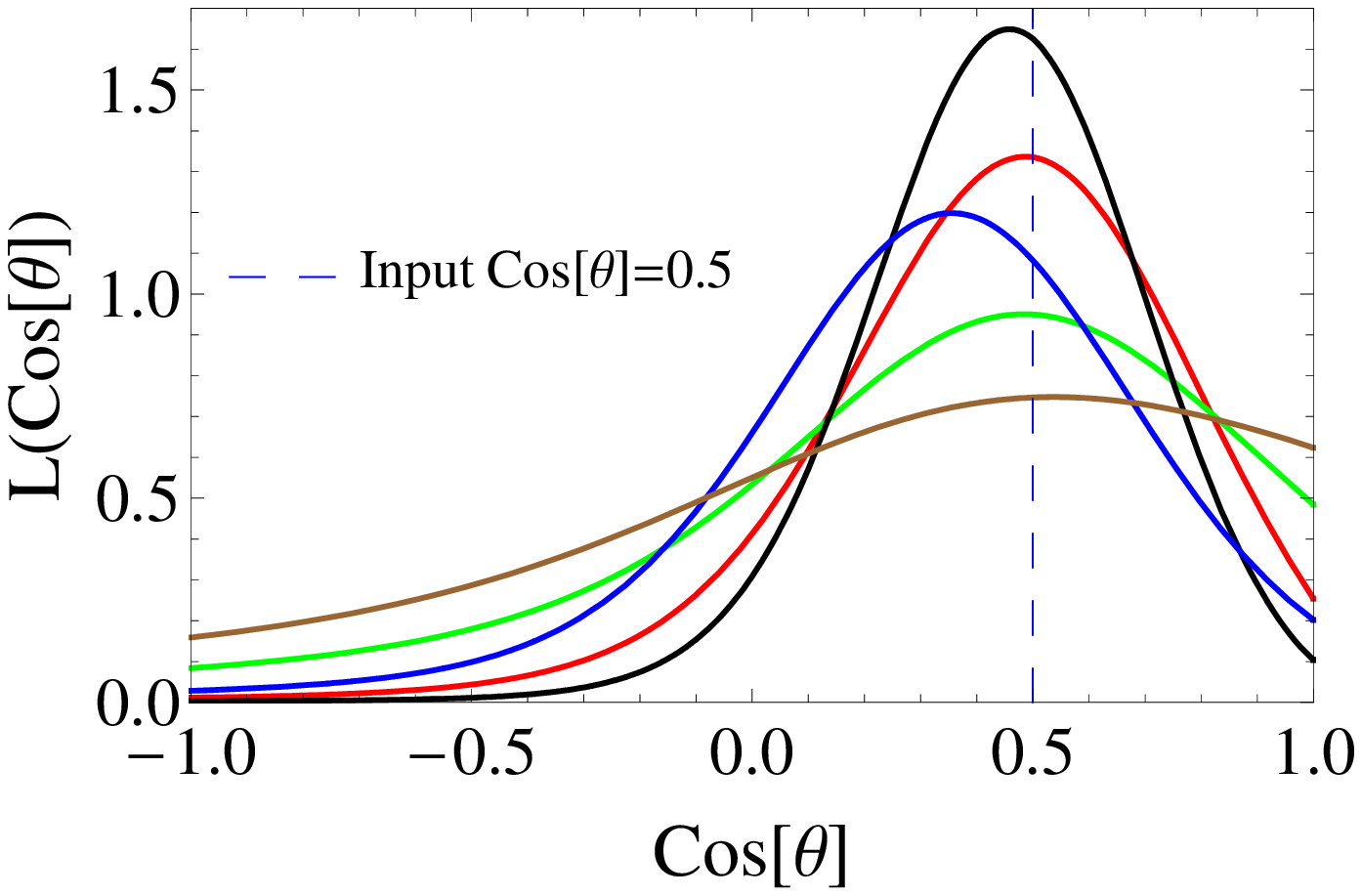}
\includegraphics[bb=0 0 430 290,width=3.05in]{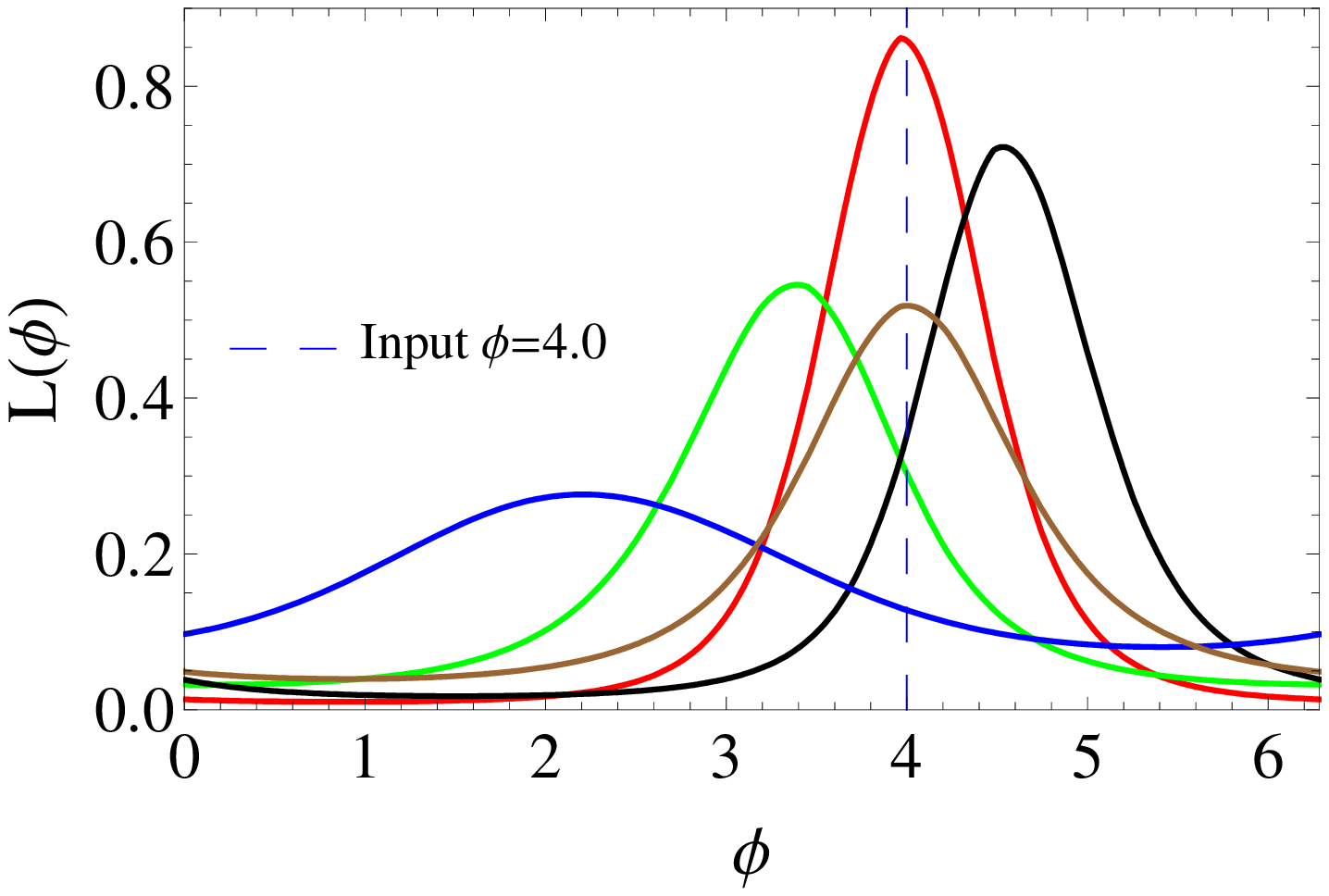}}
 \caption{A test of the likelihood function (Eqs.~(\ref{likexyz}) and
 (\ref{likev})). ($\cos(\theta)$,$\phi$) is the direction of the bulk flow
 ($\theta=\pi/2-b$,$\phi=l$). The description of this test is in Section~\ref{sec:likelihood}.}
 \label{fig:liketest}
\end{figure*}

Now, let us move on to the issue of computing the likelihood of
the magnitude and direction of the bulk flow. In general, for a
peculiar velocity survey with $N$ number of objects (galaxies,
galaxy clusters or Type Ia supernovae), of the $n$th object we can
obtain its redshift $z_{n}$, distance $r_{n}$ (utilizing the
empirical relation such as Tully-Fisher relation or the
Fundamental Plane method, see Section~\ref{sec:catalogue}), the
line of sight velocity $S_{n}$ and its measurement error $\sigma
_{n}$, and the Galactic longitude and latitude $(l,b)$
$(-90^{\circ} \leq b\leq 90^{\circ},0 \leq l\leq 360^{\circ} )$.
The line of sight component $S_{n}$ of peculiar velocity is
related to redshift ($z_{n}$) and distance ($r_{n}$) as
$cz_{n}=H_{0}r_{n}+ S_{n}$ \citep{Kaiser88,Sarkar07}.
Given the bulk motion $\mathbf{V}$ of objects in the sample, the
residual line of sight velocity of the $n$th object after
subtracting out the
bulk motion is $\delta S_n=S_{n}-\hat{r}_{n,i}V_{i}$, where $\hat{r}%
_{n,i}V_{i}$ is the projected component of $\mathbf{V}$ on to the
direction of line of sight \footnote{The three components equal to
$\hat{r}_{n,x}=\cos (b_{n})\cos (l_{n}),\hat{r}_{n,y}=\cos
(b_{n})\sin (l_{n})$ and $\hat{r}_{n,z}=\sin (b_{n})$, and $i$ is
summed over $x$,$y$,$z$ in $\hat{r}_{n,i}V_{i}$.}. After the
subtraction the residual 1D velocities should have variance
$\sigma_{n}^{2}+\sigma_{\ast }^{2}$, where $\sigma_{\ast }$
accounts for the 1D intrinsic dispersion at small scales and
$\sigma_{n}$ is the measurement error \footnote{Since
spectroscopic error of redshift is negligible, $\sigma_{n}$ is
related to the distance measurement error by
$\sigma_{n}=H_{0}*r_{n}$ (linear propagation of uncertainties).}.
Finally, the likelihood for $\mathbf{V}$=($V_{x},V_{y},V_{z}$) is
constructed as \citep{Kaiser88,Sarkar07}
\begin{equation}
L(\mathbf{V},\sigma _{\ast
})=\prod\limits_{n=1}^{N}\frac{1}{\sqrt{\sigma
_{\ast }^{2}+\sigma _{n}^{2}}}\exp \left( -\frac{1}{2}\frac{\left( S_{n}-%
\hat{r}_{n,i}V_{i}\right) ^{2}}{\left( \sigma _{\ast }^{2}+\sigma
_{n}^{2}\right) }\right)\ . \label{likexyz}
\end{equation}%
Then we transform Eq.~(\ref{likexyz}) into the spherical
coordinate system to give the joint likelihood of magnitude $V$,
direction angle $\Omega _{V}(\cos \theta ,\phi )$ and $\sigma
_{\ast }$,

\begin{widetext}
\begin{eqnarray}
&& L(V,\Omega _{V},\sigma _{\ast }) = L(\mathbf{V},\sigma _{\ast
})V^{2} \nonumber \\
&& = \prod_{n=1}^{N} \frac{V^{2}}{\sqrt{\sigma _{\ast }^{2}+\sigma
_{n}^{2}}}\exp
\left[-\frac{1}{2}\frac{1}{(\sigma_{\ast}^{2}+\sigma^{2}_{n})}
\left( S_{n}-V\left((\cos b_{n}\cos l_{n})(\sin \theta\cos \phi)
+(\cos b_{n}\sin l_{n})(\sin \theta \sin \phi) +(\sin b_{n})(\cos
\theta) \right) \right) \right].
 \label{likev}
\end{eqnarray}
\end{widetext}
To obtain the distribution of each parameter, we marginalize over
the other parameters in the likelihood function (\ref{likev}).

To assess performance of this likelihood function, we simulate 300
mock catalogues and test the behaviour of the likelihood with
these simulated data. In each mock catalogue, we simulate 100
Type-Ia supernovae data as one data set. The way we simulate each
data set is as follows. We assume that in each data set, the
supernovae share a bulk flow velocity ($V=500 \kms$) towards the
direction of ($\cos(\theta)=0.5$,$\phi=4.0$) \footnote{$\theta$ is
the angle measured from $z=0$, and $\phi$ is the azimuthal angle
measured from $x=0$. In galactic coordinate, the angles are
$l=229^{\circ}$, $b=30^{\circ}$, and we choose this arbitrary
direction just for simulation tests.}, while each line of sight
velocity has $\sigma_{\ast}=400 \kms$ random motion. We also take
the measurement error of 100 samples in the ``First Amendment
Type-Ia Supernovae'' (in abbreviation `A1SN') catalogue (see
Section~\ref{sec:catalogue}) as the measurement errors in our
simulated data sets because these quoted errors are realistic
representatives of the noise of Type-Ia supernovae. Therefore, in
each mock catalogue, we simulate the line of sight velocity of 100
Type-Ia supernovae samples, which share a streaming motion while
each has both random error and measurement error.

We then use each mock catalogue to constrain ($V, \sigma_{\ast},
\cos(\theta),\phi$) parameters, and plot the marginalized
likelihood of each parameter. The constraints from four mock
catalogues are demonstrated with green, black, blue and brown
lines in Fig.~\ref{fig:liketest}. All of these distribution
functions are centred around the input value of the parameters,
the scattering of their peak positions is determined by the number
of each sample, which scales as $1/\sqrt{N}$ with $N$ being the
number of objects in the mock. In addition, the width of each
distribution is determined by the measurement noise and intrinsic
dispersion, as they get smaller, the width of distribution becomes
narrower. We choose to simulate 100 samples in each mock catalogue
is because the simulation and the maximum-likelihood analysis can
be complete at relatively light expense of computing. Indeed,
averaging the likelihoods of 300 mock catalogues, we find that the
average distribution (red line in Fig.~\ref{fig:liketest})
perfectly peaks at the input values of preset parameters, which at
least numerically prove that the maximum-likelihood method
(Eq.~(\ref{likev})) can produce unbiased estimates of the bulk
flow.

There are other methods designed to measure bulk flow from
peculiar velocity surveys than the maximum-likelihood method. The
`All Space Constrained Estimate' (ASCE, \citealt{Nusser11}) is one
of such methods. The method is proposed in consideration of the
observational limitation that distance indicators of peculiar
velocity survey, such as Tully-Fisher relation and the Fundamental
Plane method, can only probe a small fraction of galaxies around
our local volume ($d \lesssim 100 \hmpc$) \citep{Nusser11}. To
overcome this problem, the ASCE method first generates large
number of realizations of Gaussian random velocity fields based on
the velocity power spectrum, these simulations are averaged to
obtain a series of basis functions of bulk motion. Then they fit
these basis of bulk motion with the apparent magnitude and line
width of inverse Tully-Fisher relation to estimate coefficients of
these basis. From these measured coefficients, one can therefore
reconstruct the bulk flow in our real local Universe.
\cite{Nusser11} confirmed the validity of this method with their
mock catalogues.

Another method, proposed by \cite{Branchini12}, is to use the
galaxy luminosity function at different redshifts to fit the bulk
flow velocity. Redshifts of the object may be biased by the Kaiser
rocket effect, \cite{Branchini12} provides an analytical tool to
correct this bias, and claims that it can lead to an unbiased
reconstruction of bulk flows.

We will compare our reconstruction of the bulk flow with those
found in previous studies
\citep{Watkins09,Feldman10,Nusser11,Branchini12,Turnbull12,Ma13}
in Section~\ref{sec:reconstruct}.

\section{Peculiar velocity samples and relevant treatment}
\label{sec:dataset}

\subsection{Peculiar velocity catalogues}
\label{sec:catalogue} Four different peculiar velocity catalogues
from recent surveys are adopted to estimate bulk flows. The four
samples are {\it E}arly-type {\it NEAR}by galaxies (in
abbreviation ENEAR, characteristic depth \footnote{This
characteristic depth is the `weighted-average' depth defined in
\cite{Watkins09} and \cite{Feldman10}, different from our
effective depth which considers geometry of the survey.}
$29\hmpc$, typical distance error $\sim 20$ per cent;
\citealt{Costa00,Bernardi02,Wenger03,Hudson94}), SFI++
($34\hmpc$,$\sim 23$ per cent; \citealt{Springob07}), A1SN
($58\hmpc, \sim 8$ per cent;
\citealt{Jha07,Hicken09,Folatelli10,Turnbull12}) and SC catalogue
($57\hmpc, \sim 20$ per cent; \citep{Giovanelli98,Dale99}). For
details of these samples, including characteristic depths, typical
distance errors and data compilation, please refer to section~3 of
\cite{Ma12a} and section~2 of \cite{Watkins09}.

In \cite{Feldman10} and \cite{Watkins09}, there are five other
catalogues employed, namely the SBF \citep{Tonry01}, SN
\citep{Tonry03}, SMAC \citep{Hudson99,Hudson04}, EFAR
\citep{Colless01} and Willick sample \citep{Willick99}. Here we
opt to abandon these five catalogues, and the reasons are as
follows. For SMAC, EFAR and Willick, these samples are either very
distant, in which case the distance errors are very large, or too
sparse to support robust estimation, and their survey geometry is
so complicated that make it hard to measure. In addition, as the
survey goes deeper, the simple model of assuming Gaussian errors
of distances is almost certainly inappropriate, and will become a
dominant systematic effect in the distance estimation; velocity
data beyond $100 \hmpc$ are thus too noisy to reliably reconstruct
bulk flow. For SBF data, it is too close to our own galaxy, some
galxies fall into our local non-linear structures, therefore it
could strongly bias our estimation of bulk velocity on large
scales. Since we will use the newly compiled A1SN catalogue (see
\citealt{Turnbull12} and \citealt{Ma12a}) which includes three
Type-Ia supernovae data sets, we will not use its old sub data
set, the SN set \citep{Tonry03}, in our study.

\subsection{Malmquist bias correction}
\label{sec:MBcorrect}
In the catalogues described above, there are three different
classes of distance indicators, the Tully-Fisher relation (SFI++,
SC), the Fundamental Plane method (ENEAR) and the Type-Ia SN
luminosity function (A1SN). These distance indicators all have
their intrinsic errors. For Tully-Fisher selected samples, such as
SFI++ and SC, their distance errors are around $23$ per cent,
which is slightly larger than the distance error of the
Fundamental Plane-selected ENEAR sample ($\sim 18$ per cent). For
Type-Ia supernovae data, the luminosity function can be used to
calibrate the distance in better precision, the distance errors of
A1SN catalogue are only $\sim 7$ per cent. The uncertainty of
distance indicators, especially for Tully-Fisher and Fundamental
Plane-selected objects, suggests that an object with its measured
distance $d$ may actually deviate from its true distance by a
broad range of possible values. This is the effect of Malmquist
bias \citep{Malmquist20}, which characterizes the fact that
inhomogeneous distribution of matter and distance (or magnitude)
errors can in general bias the distance (magnitude) measurement.
As a result, the probability function of the true distance $r$
given the measured distance $d$ strongly depends on the intrinsic
errors of distance indicators, and the underlying density
distribution \citep{Malmquist20,Lynden-Bell88}. Taking the
\textit{IRAS}-PSC$z$ (Point Source Catalogue with redshift)
catalogue which probes the full-sky underlying density field out
to $192 \hmpc$ as the model of cosmic matter distribution, we
follow the guideline in section 3.1 of \cite{Ma12a} and section
2.3 of \cite{Ma13} to correct Malmquist bias for A1SN, SC and
ENEAR catalogues. Note that the SFI++ catalogue \citep{Springob07}
is already corrected for Malmquist bias.

Once the Malmquist bias is corrected, our next step is to select
samples. In the four catalogues, objects with distance beyond
$100\hmpc$ are very sparse and suffer from large errors due to
uncertainties in the distance indicators, which are consequently
discarded from the sample. Additionally, several SFI++ galaxies
with $d \lesssim 30\hmpc$ are strongly affected by local
non-linear structures, showing very large velocities
\citep{Ma12a}, we also excluded these high-velocity members
($|v|>3000 \kms$) from the SFI++ catalogue since they are clearly
close to some local non-linear structures. Our final samples for
the maximum-likelihood analysis are listed  in
Table~\ref{tab:tab1}.

\begin{table}
\begin{centering}
\begin{tabular}{@{}llllll}
\hline
catalogues & $d \leq 100$ & $ d > 100$ & $R_{\rm min}$ & $R_{\rm max}$ & $b_{\rm cut}$\\
\hline
\ ENEAR & $690$ & $ 7$ & $6$ & $100$ & $3.6$ \\
\ A1SN & $175$ & $100$ & $5$ & $100$ & $0$ \\
\ SFI++ & $2915$ & $541$ & $0$ & $100$ & $8$ \\
\ SC & $28$ & $42$ & $13$ & $90$ & $18$ \\
\hline
\end{tabular}%
\caption{Final samples for analysis extracted from the four
peculiar velocity catalogues. The first two columns give the
number of galaxies within the range $d \leq 100\hmpc$ (used in
this paper) and $d>100 \hmpc$ (considered as outliers). The last
three columns delimit geometry of samples after Malmquist bias
correction and selection, $R_{\rm{min}}$ and $R_{\rm{max}}$ are
the minimal and maximal distances of samples in unit of $\hmpc$,
$|b|\le b_{\rm{cut}}$ defines the cut-off region in galactic
latitude for each survey.} \label{tab:tab1}
\end{centering}
\end{table}

\subsection{Geometry of the survey}
\label{sec:geometry}

\begin{figure*}
\centerline{\includegraphics[bb=13 175 598
616,width=3.2in]{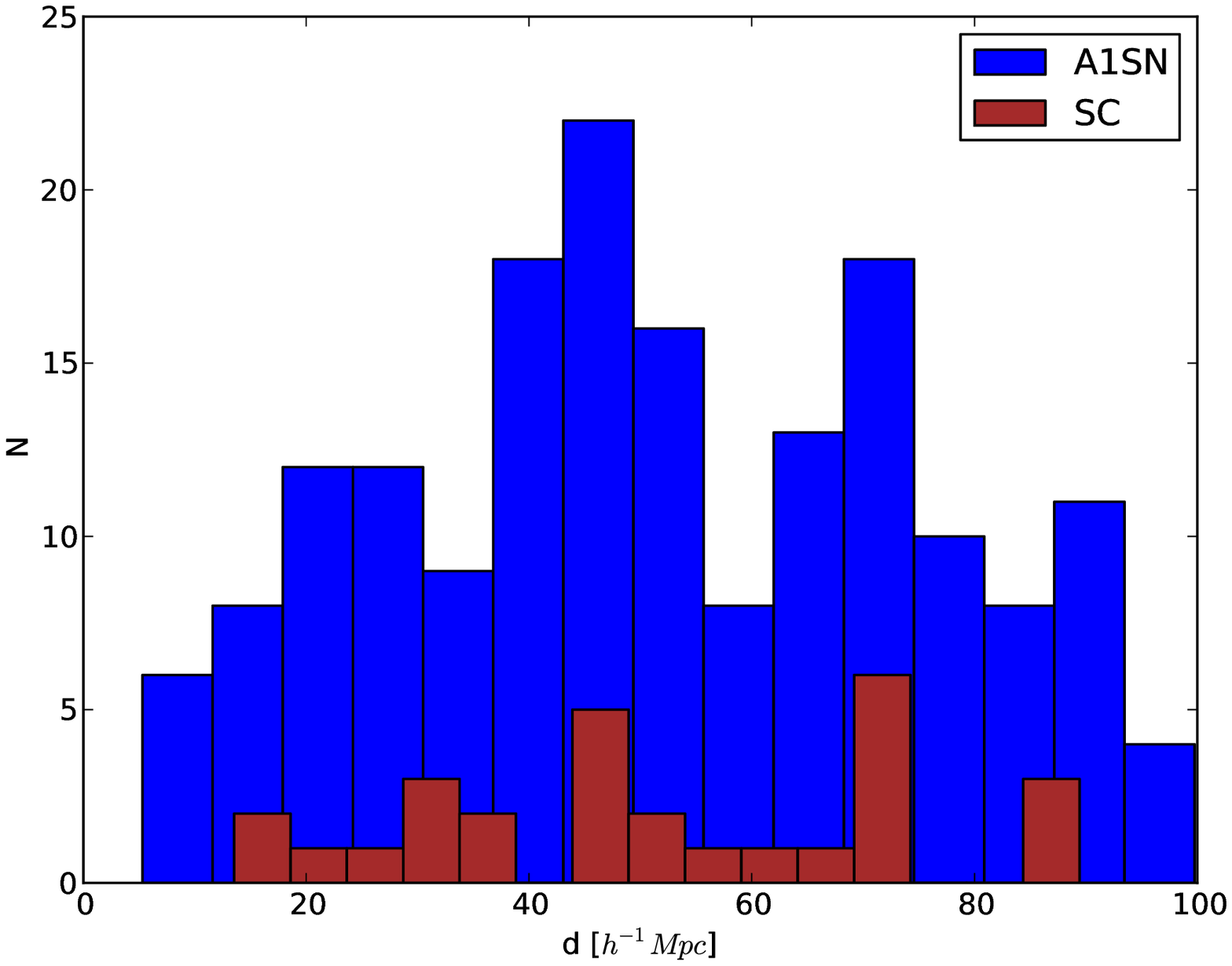}
\includegraphics[bb=13 175 598
616,width=3.2in]{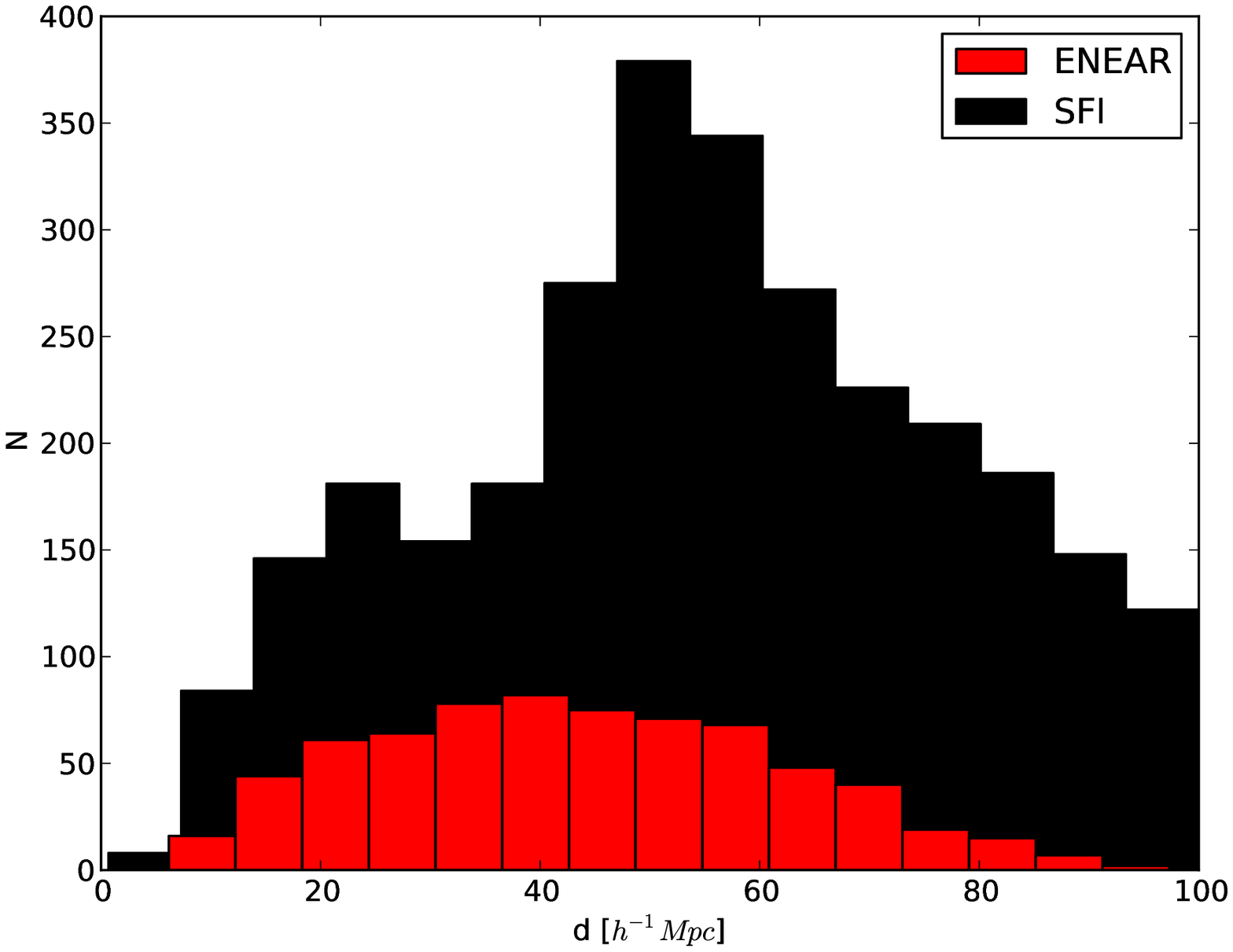}}
\centerline{\includegraphics[bb=0 0 590
300,width=3.1in]{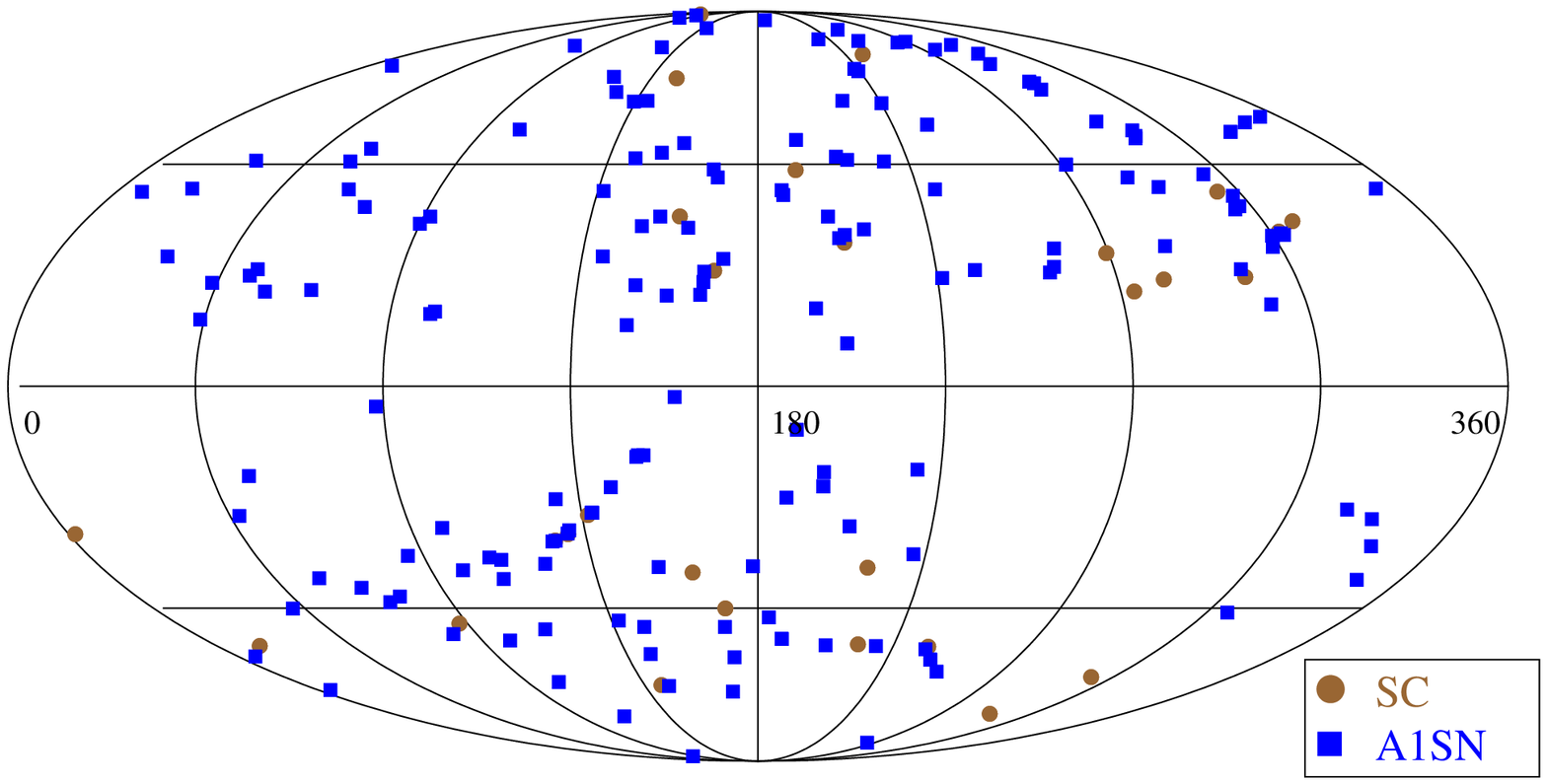}
\includegraphics[bb=0 0 590 300,width=3.1in]{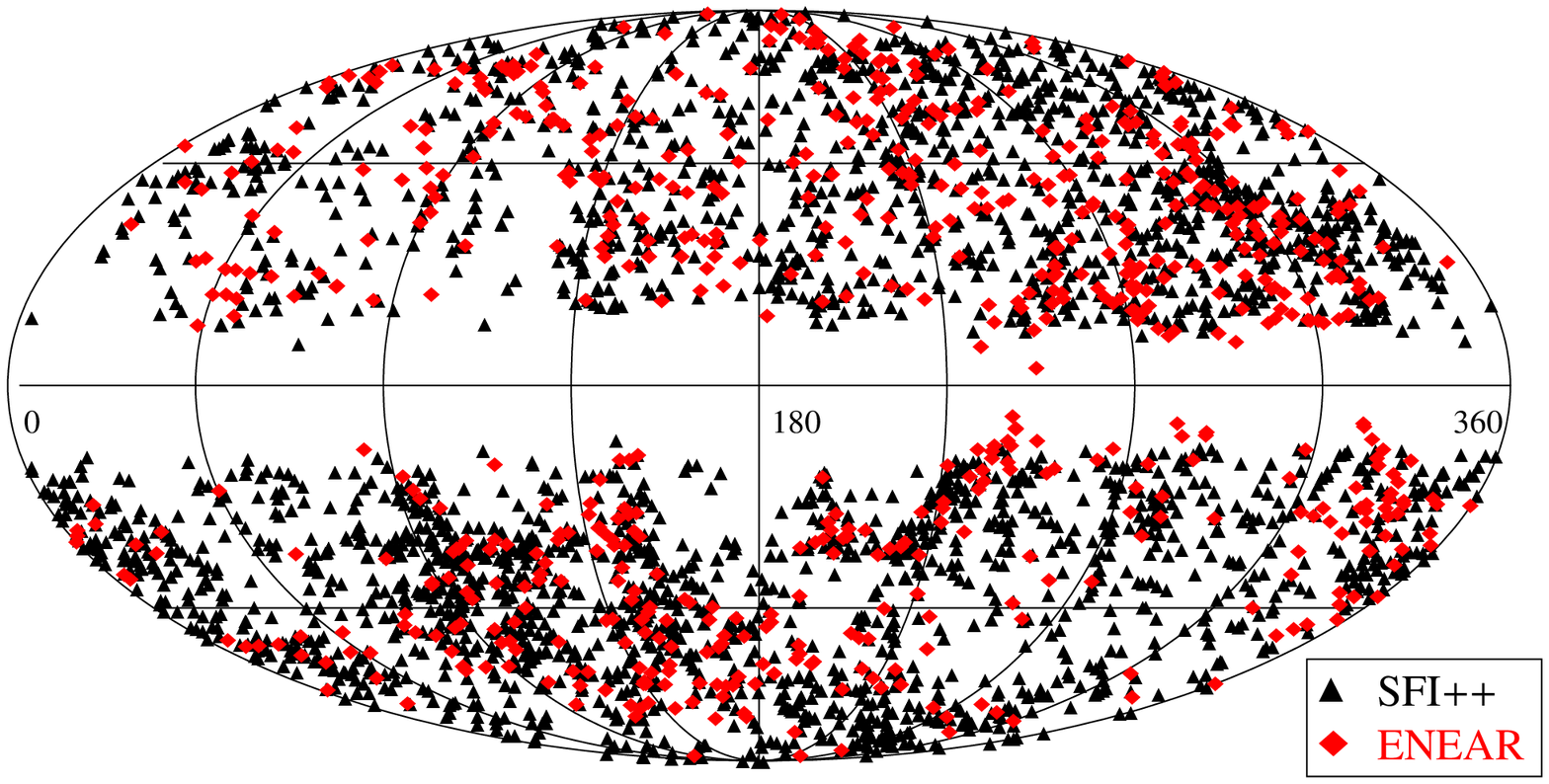}}
 \caption{Upper two panels: the distance histogram of the four catalogues; lower two panels:
 the distribution of the ENEAR, A1SN, SFI++ and SC samples on the sky. The data are quite
homogeneous across the full sky therefore robust tests of local
anisotropy can be made.}
 \label{fig:lbdist}
\end{figure*}

In Fig.~\ref{fig:lbdist} we show the histogram of the distances
for each sample along the radial direction (upper two panels), and
spatial distribution of the four samples on the sky (lower two
panels). From the upper two panels of Fig.~\ref{fig:lbdist}, one
can see that the four samples have different distance histogram,
the ENEAR catalogue is effectively the shallowest sample with the
median distance around $40\hmpc$, while the A1SN and SFI++
catalogues all have median distance around $50\hmpc$. The lower
two panels show that the four catalogues have nearly coverage the
full sky, except for the small blank region along the galactic
plane. Geometry information of the four peculiar velocity
catalogues are tabulated in Table~\ref{tab:tab1}.

Now let us turn into the issue of calculating characteristic depth
of each catalogue. The galaxy peculiar velocity survey can probe
only limited depth with full or partial sky coverage. Therefore,
the characteristic depth of the samples are strongly affected by
this effective survey volume. \cite{Ma13} and \cite{Turnbull12}
calculated the effective depth as the average of distances of all
member objects. They weighted the distance of every object with
the square of the inverse of its distance error, i.e.
\begin{eqnarray}
\overline{r}=\frac{\sum_{n}r_{n}/\sigma^{2}_{n}}{\sum_{n}1/\sigma^{2}_{n}}\
. \label{inverseweight}
\end{eqnarray}

However, the weighted-average distance does not take into account
the radial distribution of the survey, as well as the influence of
partial sky coverage. In this work, we adopt an alternative
approach for the characteristic depth calculation proposed by
\cite{Li12} to take care of these effects. Considering the real
survey geometry (Table~\ref{tab:tab1}) and the radial distribution
function, we identify that the `true' survey window function is
$W_{\rm{true}}(\mathbf{x})=W(\mathbf{x})n(\mathbf{x})$, where
$n(\mathbf{x})$ is the 3D density distribution. Fourier
transforming the `true' window function gives
\begin{eqnarray}
\tilde{W}(k)=\int W_{\rm{true}}(\mathbf{x}) {\rm e}^{-{\rm i}
\mathbf{k} \cdot \mathbf{x}} \der ^{3} \mathbf{x}\ ,
\label{fourierwin}
\end{eqnarray}
which can be plugged into Eq.~(\ref{sigmav}) to yield an effective
rms of bulk flow velocity (detailed calculation is in Appendix
\ref{filterwin})
\begin{eqnarray}
\tilde{V}_{\rm{rms}}^{2}(R) = \frac{1}{(2\pi )^{3}}\int P_{\rm
vv}(k)\tilde{W}^{2}(k) \der^{3}k\ . \label{sigmav3}
\end{eqnarray}%
The value of this velocity rms is the expectation of linear theory
for the true window function. The effective depth of the sample is
defined as the radius $R$ of the top-hat window function which
offers the same theoretical velocity rms (Eq.~(\ref{sigmav2})) as
the true window function does. The effective depth is so in the
sense that it filters the same modes of perturbation as the true
survey window function. We list our findings of effective depth in
the first column of Table~\ref{tab2}. This characteristic depth
will be used to locate the position of the bulk flow magnitude on
the velocity--distance diagram (Fig.~\ref{fig:Constraints}).

\section{Results}
\label{sec:results}

\subsection{Reconstructed bulk flow}
\label{sec:reconstruct}
\begin{figure*}
\centerline{\includegraphics[bb=0 0 490
329,width=2.2in]{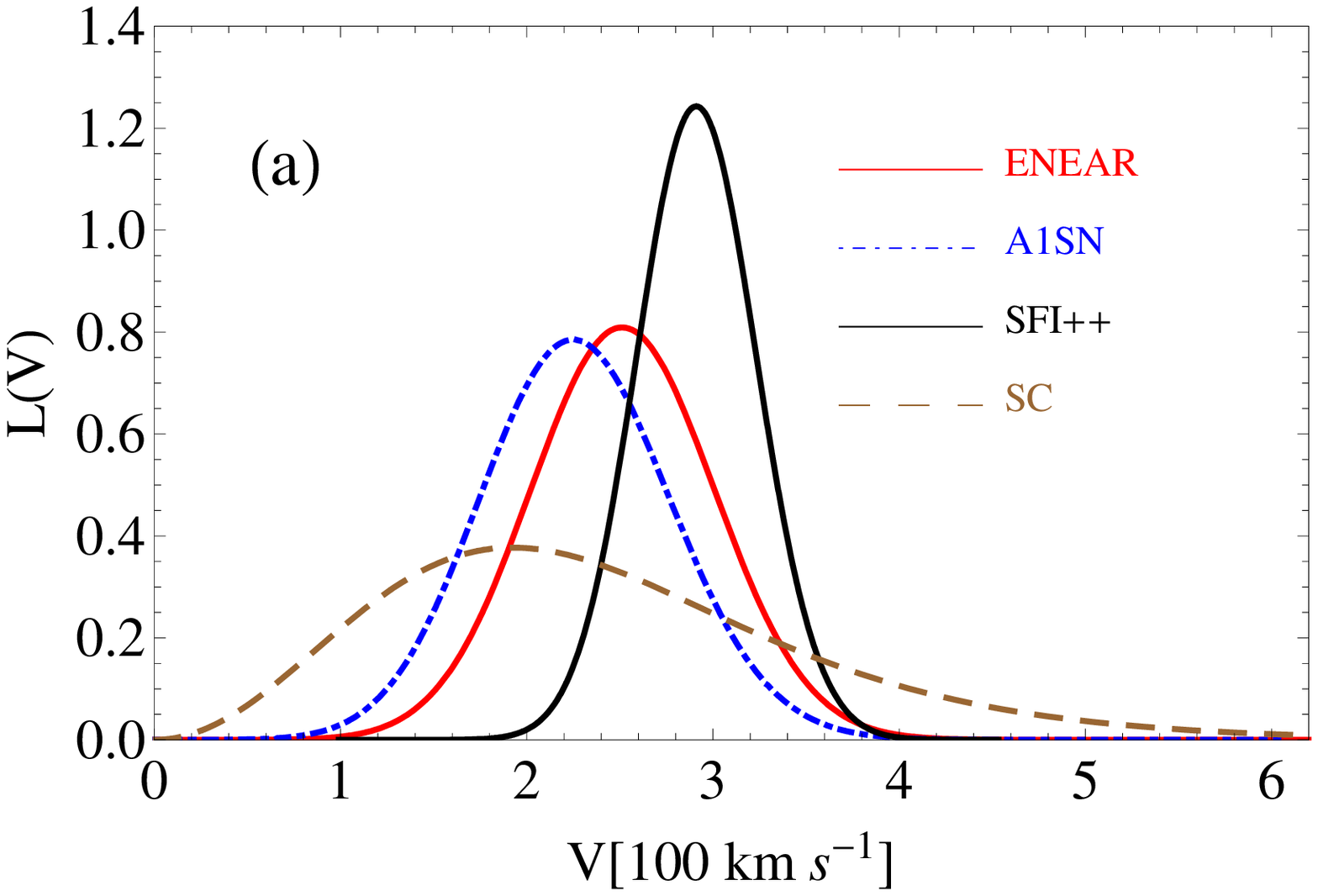}
\includegraphics[bb=0 0 497 329,width=2.2in]{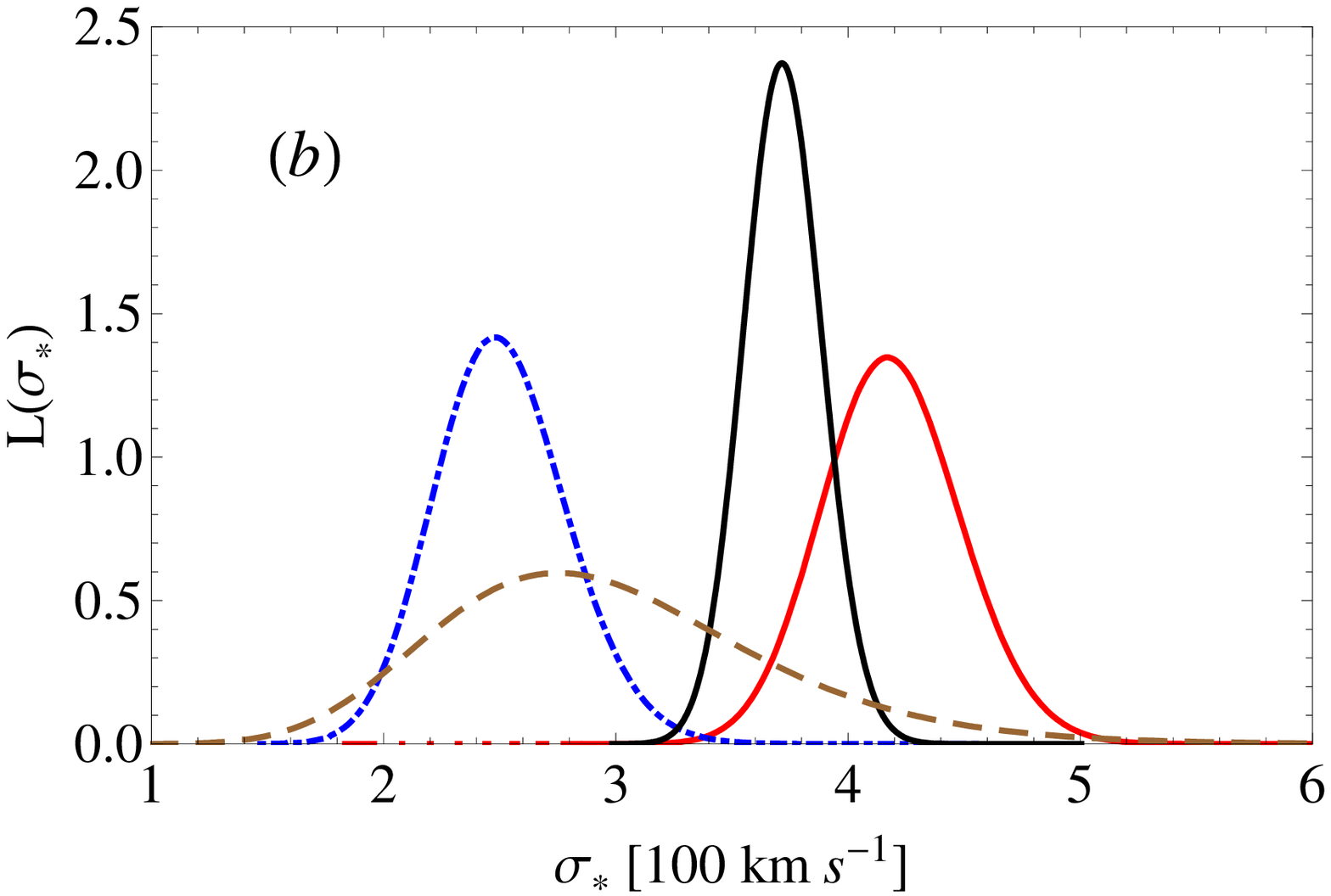}
\includegraphics[bb=0 0 527 530,width=2.2in]{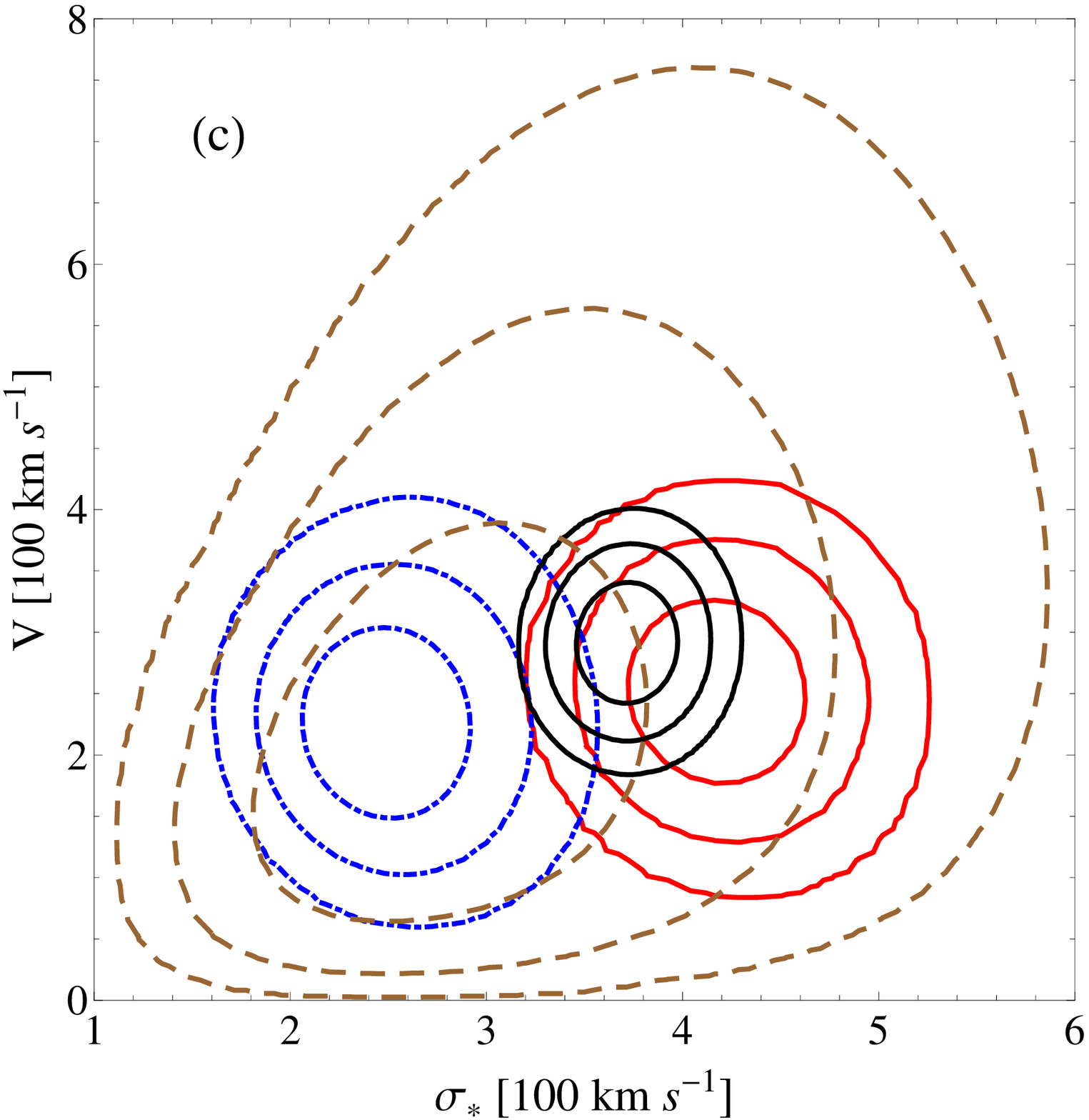}}
\centerline{\includegraphics[bb=0 0 486
326,width=2.2in]{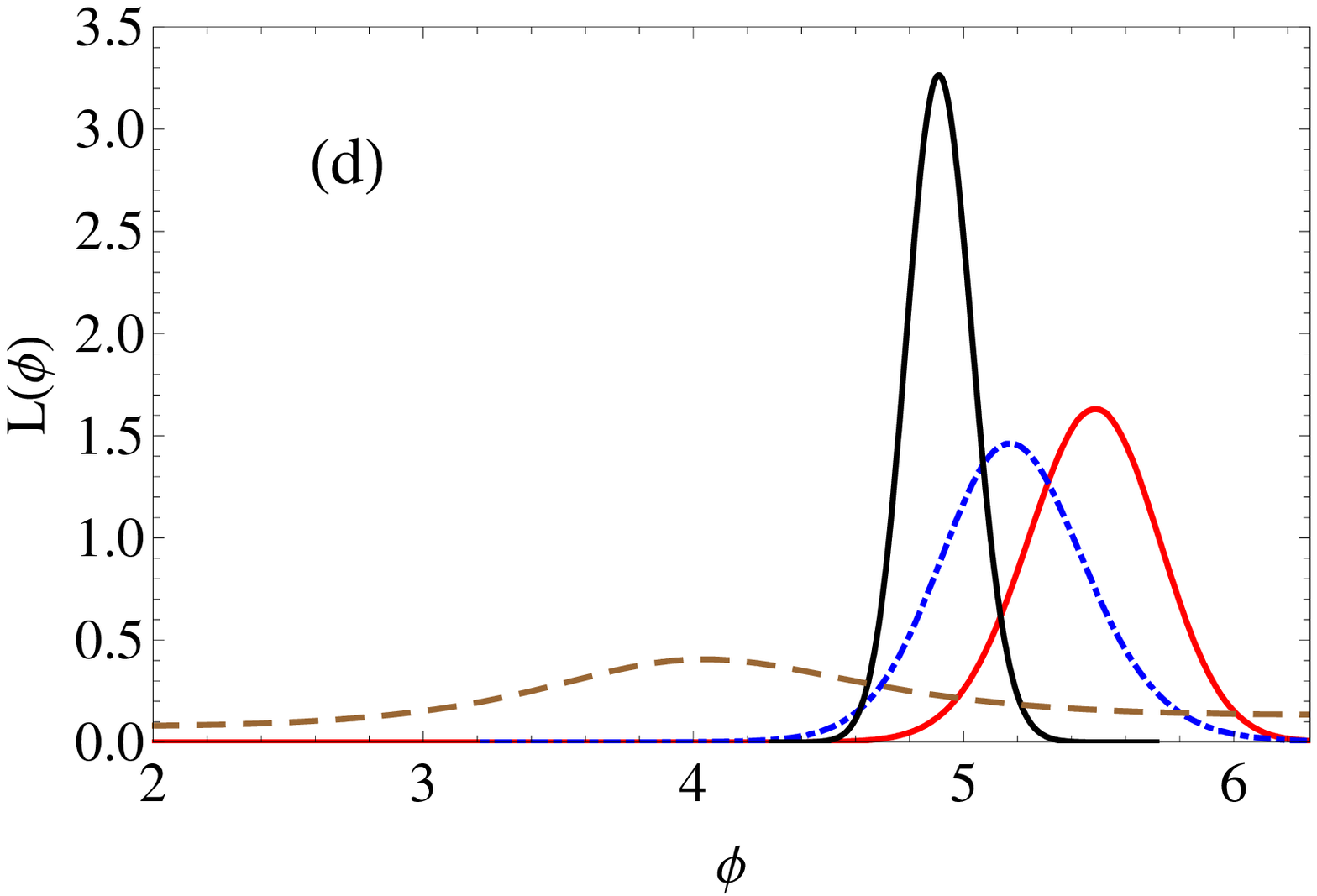}
\includegraphics[bb=0 0 500
343,width=2.2in]{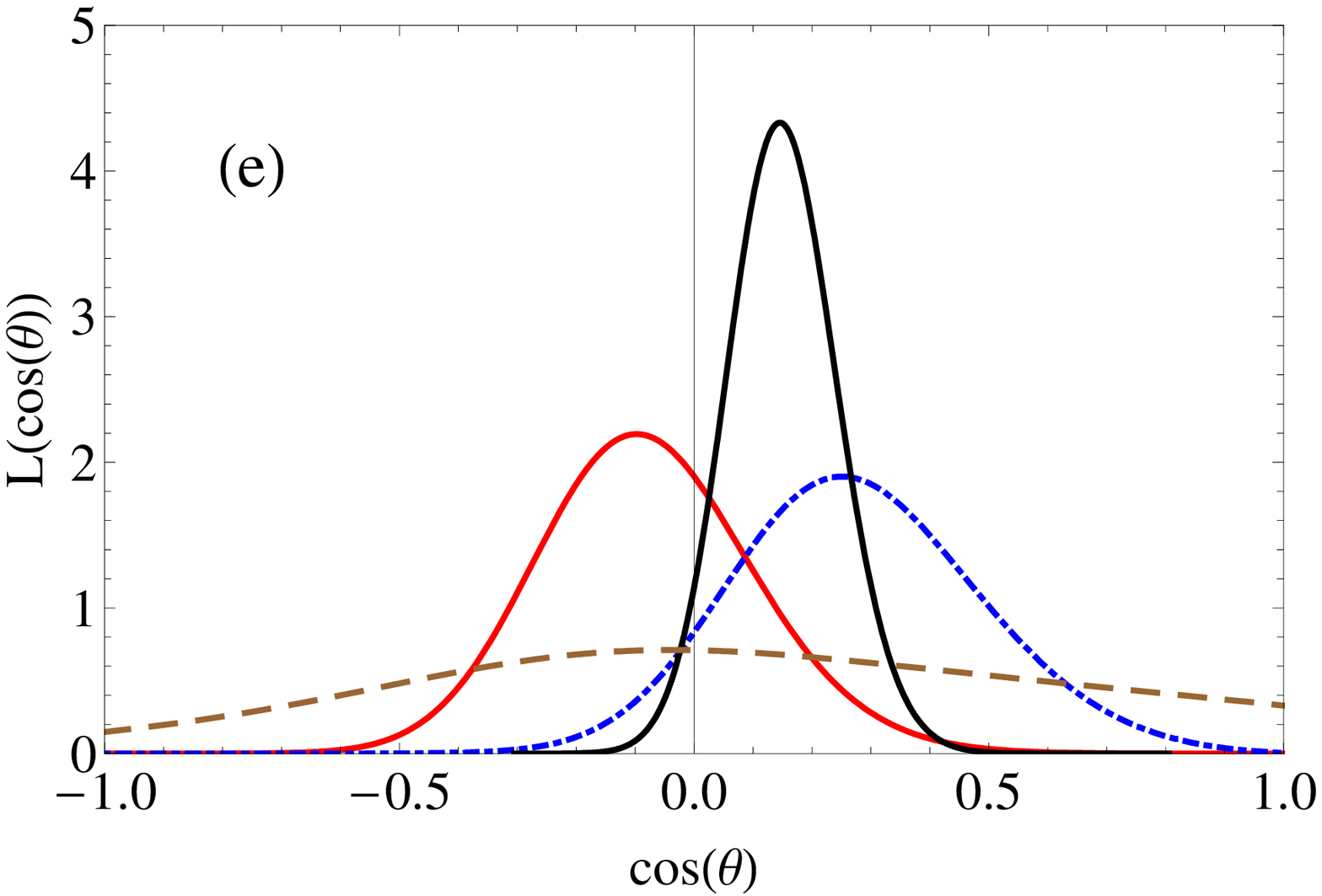}
\includegraphics[bb=0 0 494 494,width=2.2in]{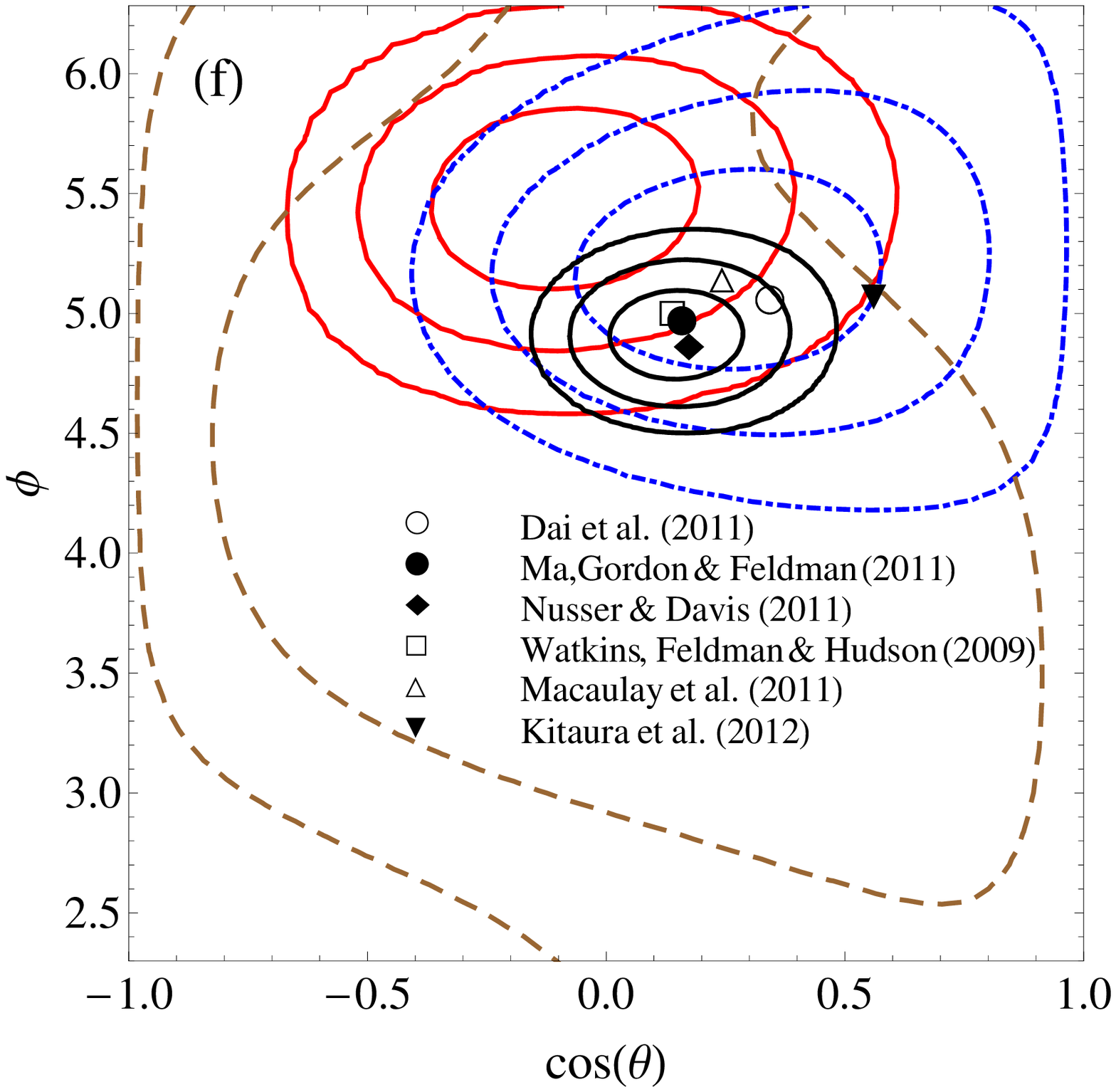}}
 \caption{Marginalized likelihood function of magnitude $V$, direction ($\cos(\theta)$,$\phi$) of the
 bulk flow, and the small-scale dispersion
 $\sigma_{\ast}$. (We marginalize Eq.~(\ref{likev}) with respect to other parameters
 to obtain the likelihood of parameters of interest.) We use $\theta=\pi/2-b$,$\phi=l$, since
 ($\cos(\theta)$,$\phi$) are Gaussian-distributed variables. Panel (c)
 shows that different catalogues prefer different values of
 $\sigma_{\ast}$, but they all provide consistent constraints on
 bulk flow magnitude $V$.
 From panel (f), one can see that the bulk flow directions
 reconstructed by us are consistent with other works.}
 \label{fig:like}
\end{figure*}

The likelihood function (Eqs.~(\ref{likexyz}) and (\ref{likev}))
is applied for estimation of
($V$,$\cos(\theta),\phi$,$\sigma_{\ast}$) to the four peculiar
velocity samples. In Fig.~\ref{fig:like}, we plot the constraints
on the bulk flow $\mathbf{V}$ (panels a, d and e), the small-scale
dispersion $\sigma_{\ast}$ (panel b), and joint likelihood
contours on planes of ($V$,$\sigma_{\ast}$) and
($\cos(\theta)$,$\phi$)  (panels c and f).

From Fig.~\ref{fig:like}a, one can see that since different
surveys probe different volumes of the Universe, peaks of the
likelihood functions locate at different values, reasonable
comparison ought to be made together by their characteristic
depths. Comparison of the constraints with the theoretical model
is in Section~\ref{sec:compare}.

In Fig.~\ref{fig:like}b, we plot the likelihood of the small-scale
intrinsic velocity dispersion. It is apparent that each catalogue
prefers different $\sigma_{\ast}$. For the Type-Ia supernovae
sample (A1SN) and the SC catalogue, $\sigma_{\ast}$ is around $250
\kms$, but for the SFI++ and the ENEAR, $\sigma_{\ast}$ is around
$400 \kms$. The value of $\sigma_{\ast}$ reflects the disturbance
on very small scales, whilst bulk motion reflects perturbation on
large scales. Thus, the bulk motion $V$ and the $\sigma_{\ast}$
should not correlate with each other, which is verified by the
(nearly) orthogonal contours shown in Fig.~\ref{fig:like}c.

We further plot the likelihood of the direction angle
$\cos(\theta)$ (Fig.~\ref{fig:like}e) and $\phi$
(Fig.~\ref{fig:like}d), and their correlation contours
(Fig.~\ref{fig:like}f). By comparing the ($\cos(\theta)$,$\phi$)
contours in Fig.~\ref{fig:like}f, with the direction angle probes
by the previous studies, we can find that the direction angles
constrained from our A1SN, ENEAR and SFI++ catalogues are pretty
well consistent with the Type-Ia supernovae constraints by
\cite{Dai11}, the SFI++ constraints by \cite{Nusser11}, the
combined catalogue constraints by \cite{Watkins09} and
\cite{Ma11}, and the reconstructed Two-Micron All-Sky Redshift
Survey density field \cite{Kitaura12}.

We list the results of our constraints in Table~\ref{tab2}.
Comparison of our results with other reconstructed bulk flow of
the top-hat window function is in Table~\ref{tab3}. In comparison,
we also list the reconstructed bulk flow of Gaussian window
function in Table~\ref{tab4}.

\begin{table*}
\begin{centering}
\begin{tabular}{@{}llllll} 
\hline
Catalogues & $R (\hmpc)$ & $V$ ($100 \kms$) & $\sigma_{\ast}$ ($100 \kms$) & $l$ ($^{\circ}$) & $b$ ($^{\circ}$)  \\
\hline \noalign{\vspace{1pt}}
\ ENEAR & $49$ & $2.5 \pm 0.5$ & $4.2 \pm 0.3$ & $314 \pm 14$ & $-6^{+11}_{-9}$ \\
\ SFI++ & $58$ & $2.9 \pm 0.3$ & $3.7 \pm 0.2$ & $281 \pm 7$ & $8^{+6}_{-5}$ \\
\ A1SN & $62$ & $2.3 \pm 0.5$ & $2.5 \pm 0.3$ & $296 \pm 16$ & $15^{+13}_{-12}$ \\
\ SC  & $63$ & $1.9^{+1.2}_{-0.9}$ & $2.8^{+0.8}_{-0.6}$ & $231^{\times}_{\times}$ & $-2^{+35}_{-31}$ \\
\hline
\end{tabular}%
\caption{The results of constraints from four catalogues. The $R$
is the effective top-hat window size of the sample after the
Malmquist bias correct and sample selection. The error bars listed
are quoted for $1\sigma$ confidence level (CL).} \label{tab2}
\end{centering}
\end{table*}

\begin{table*}
\begin{centering}
\begin{tabular}{@{}llllllll} 
\hline
Samples & Numbers & $R (\hmpc)$  & $V$ ($100 \kms$) & $l$ ($^{\circ}$) & $b$ ($^{\circ}$) & Method & References  \\
\hline \noalign{\vspace{1pt}}
\ SFI++ & $2915$ & $58$  & $2.9 \pm 0.3$ & $281 \pm 7$ & $8^{+6}_{-5}$ & MLE & This study \\
\ SFI++ & $2895$ & $40.0$  & $3.3 \pm 0.4$ & $276 \pm 3$ & $14 \pm 3$ & ASCE & \cite{Nusser11} \\
\ SFI++ & $2895$ & $100.0$  & $2.6 \pm 0.4$ & $279 \pm 6$ & $10 \pm 6$ & ASCE & \cite{Nusser11} \\
\hline
\ ENEAR & $690$ & $49$ & $2.5 \pm 0.5$ & $314 \pm 14$ & $-6^{+11}_{-9}$ & MLE & This study \\
\hline
\ A1SN & $175$ & $62$ & $2.3 \pm 0.5$ & $296 \pm 16$ & $15^{+13}_{-12}$ & MLE & This study \\
\ SN & $133$ & $45.0$ & $2.8 \pm 0.7$ & $285 \pm 18$ & $-10 \pm 15$ & MCVF & \cite{Haugboelle07} \\
 \hline
\end{tabular}%
\caption{Comparison of the reconstructed bulk flow of the top-hat
window function with the studies in the literature. The error bars
listed are for $1\sigma$ CL. `MLE' stands for `Maximum-Likelihood
Estimate'; `ASCE' stands for `All Space Constrained Estimate';
`MCVF' stands for the method of extracting Multipole Components of
the Velocity Field. The sample used in \citet{Haugboelle07} is the
Type-Ia supernovae data from \citet{Hicken09} and \citet{Jha07}
respectively.} \label{tab3}
\end{centering}
\end{table*}

\begin{table*}
\begin{centering}
\begin{tabular}{@{}llllllll} 
\hline
Samples & Numbers & $R (\hmpc)$  & $V$ ($100 \kms$) & $l$ ($^{\circ}$) & $b$ ($^{\circ}$) & Method & References  \\
\hline \noalign{\vspace{1pt}}
\ SFI++ & $2404$ & $50.0$ & $3.4 \pm 0.4$ & $280 \pm 8$ & $5.1 \pm 6$ & MV & \cite{Ma13} \\
\ SFI++ & $3401$ & $34.0$ & $4.3 \pm 1.0$ & N/A & N/A & MV & \cite{Watkins09} \\
\hline
\ ENEAR & $669$ & $50.0$ & $2.2 \pm 0.6$ & $310 \pm 30$ & $-9.8 \pm 14$ & MV & \cite{Ma13} \\
\hline
\ A1SN & $153$ & $50.0$ & $2.2 \pm 0.7$ & $290 \pm 60$ & $12.1 \pm 60$ & MV & \cite{Ma13} \\
\ A1SN & $245$ & $58.0$ & $2.5 \pm 0.7$ & $319 \pm 18$ & $7 \pm 14$ & MV & \cite{Turnbull12} \\
\ SN & $557$ & $150.0$ & $1.9^{+1.2}_{-1.0}$ & $290^{+39}_{-31}$ & $20 \pm 32$ & MCMC & \cite{Dai11} \\
\ SN & $112$ & $40.0$ & $5.4 \pm 0.9$ & $258 \pm 10$ & $36 \pm 11$ & WLS & \cite{Weyant11} \\
\ SN & $112$ & $40.0$ & $4.5 \pm 1.0$ & $273 \pm 11$ & $46 \pm 8$ & CU & \cite{Weyant11} \\
\hline
\ COMPO & $4356$ & $50.0$ & $4.1 \pm 0.8$ & $287 \pm 9$ & $8 \pm 6$ & MV & \cite{Watkins09} \\
 \hline
\end{tabular}%
\caption{Comparison of the reconstructed bulk flow of the Gaussian
window function with the studies in the literature. The error bars
listed are for $1\sigma$ CL. `MLE' stands for `Maximum-likelihood
Estimate'; `MV' stands for the `Minimal Variance' method; `MCMC'
stands for the Bayesian Markov Chain Monte Carlo method; `WLS'
stands for `weighted least squares'; `CU' stands for the
`coefficient unbiased' method. The sample used in \citet{Dai11} is
the Union2 catalogue of Type-Ia supernovae \citep{Amanullah10},
and the samples used in \citet{Weyant11} are the Type-Ia
supernovae data from \citet{Hicken09} and \citet{Jha07}
respectively. The final row shows the result of COMPOSITE data
set, which is a combined catalogue from eight different peculiar
velocity surveys (SBF, ENEAR, SFI++, SN, SC, SMAC, EFAR,
Willick).} \label{tab4}
\end{centering}
\end{table*}



\subsection{Comparing with theoretical prediction}
\label{sec:compare}

\begin{figure}
\centerline{\includegraphics[bb=11 8 593
404,width=3.0in]{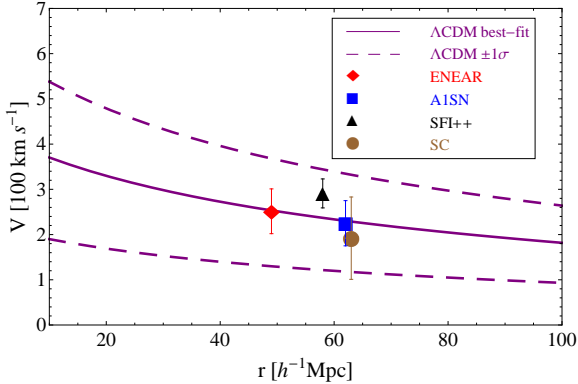}}
 \caption{The comparison between the velocity magnitude from the likelihood function (Eq.~(\ref{likev})), and
 the theoretical evaluation (Eq.~(\ref{pveq})) given different depths of survey.}
 \label{fig:Constraints}
\end{figure}

We plot our results of the constraints (Table~\ref{tab2}) in
Fig.~\ref{fig:Constraints} together with the predictions of the
$\Lambda$CDM model. The solid line is the peak ($V_{\rm p}$) of
the distribution (Eq.~(\ref{pveq})), and the dashed lines are the
$\pm 1\sigma$ confidence interval. One can see that the data are
consistent with the expectation for the $\Lambda$CDM cosmology.
Note that the SFI++ catalogue, with its nearly full-sky coverage
and dense sampling, provides the tightest constraint of the bulk
flow amplitude. In addition, by comparing our constraints with the
other studies in Table 1 and the earlier works
\citep{Courteau93,Willick97,Courteau00} by using Mark III and
Shallflow catalogues, we can see that they all provide constraints
on bulk flow amplitude ($ \sim 300 \kms$) on scales of $50\hmpc$
that are consistent with the prediction for the $\Lambda$CDM
model.

If the data are improved by prospective new surveys, such as 6dF
survey \citep{Jones09} or Square Kilometer Array \citep{ska}, the
data can be used to constrain any possible deviation from general
relativity, i.e. the standard gravity theory. This is because any
alternative theory of gravity would change the growth rate of
structure, which will boost or diminish the power of the velocity
field on intermediate scales ($0.01 \mpch \lesssim k \lesssim 100
\mpch$) (see fig.~2 in \citealt{Ma12b}). While doing the
constraints on the modified gravity model, one needs to keep in
mind of the sample variance at different scales, which is plotted
as the dashed line in Fig.~\ref{fig:Constraints}. For each scale
$R$, there is a certain level of uncertainty of fluctuations which
reflects the variation of the number of velocity modes filtered by
the window function. The sample variance limits the capacity of
the reconstructed bulk flow to infer the underlying physics, one
needs to consider this variance term in the full covariance matrix
when the bulk flow is used to constrain cosmology.

\section{Conclusion}
\label{sec:conclude}

As introduced in Section~\ref{intro}, bulk flow is the coherent
motion of sampled galaxies, galaxy clusters or supernovae, which
can be used as a test of the growth of structure. Yet there are
some tentative observational evidences from the peculiar velocity
surveys and the CMB observation suggesting possible excess power
on scales around $50 \hmpc$ (in radius). Here, we show that data
of current peculiar velocity surveys actually do not provide
strong evidence against the $\Lambda$CDM model.

In this paper we adopted a maximum-likelihood method to peculiar
velocity catalogues for the bulk flow estimation. Different from
just using the `peak' of the maximum-likelihood method as in
\cite{Kaiser88}, we employ the full likelihood function with
simulated data sets and the state-of-the-art peculiar velocity
survey. Numerical test with simulations indicates that the
estimator is unbiased in the limit that no complicated survey
geometry is involved, which is approximately true for the four
catalogues.

We apply our likelihood function to the four catalogues, ENEAR,
SFI++, A1SN and SC all of which are Malmquist bias corrected and
properly trimmed, to obtain the magnitude and direction of the
bulk flows. We find: (1) for the largest and densest Tully-Fisher
selected catalogue, SFI++ survey constrain the magnitude of bulk
flow as $V=290 \pm 30 \kms$ towards $l=281^{\circ} \pm 7^{\circ}$
at an effective depth of $58\hmpc$,
$b=8^{\circ+6^{\circ}}_{-5^{\circ}}$; (2) for the largest
Fundamental Plane selected catalogue, ENEAR samples constrain the
bulk flow as $V=250 \pm 50 \kms$ towards $l=314^{\circ} \pm
14^{\circ}$, $b=-6^{\circ +11^{\circ}}_{-9^{\circ}}$ at an
effective depth $49\hmpc$. Directions of the bulk flow we find
here are well consistent with the previous probes, while
amplitudes of estimated bulk flows confirm an earlier
investigation with the same data sets but different estimation
method \citep{Ma13}.

From the geometry of the selected peculiar velocity samples,
incorporating the radial distribution of sampled objects, we
calculate the effective depth for the four velocity samples. Our
estimated bulk speeds are placed together with the theoretical
prediction of the flow at the effective depths we computed, and we
find that the two matches pretty well on all scales till $50
\hmpc$. The results of this paper clearly show that, the bulk flow
velocities constrained from currently available peculiar velocity
surveys do not demonstrate sign of excess large amplitudes, but
rather are in full agreement with the gravitationally induced bulk
flow as predicted by the concordance model of $\Lambda$CDM
cosmology.

\vskip 0.1 truein

\noindent \textbf{Acknowledgements:} We would like to thank the
discussion with Neta Bahcall, Anthony Challinor, William Saslaw,
Jasper Wall and Gongbo Zhao and the two anomalous referees. YZM is
supported by a CITA National Fellowship. This research is
supported by the Natural Science and Engineering Research Council
of Canada.

\appendix

\section{The filtered window function}
\label{filterwin} In Section~\ref{sec:geometry}, we want to take
the specific survey volume into account and calculate its real
window function and rms of the velocity. We identify the true
survey window function as
\begin{equation}
W_{\rm{true}}(\mathbf{x})=W(\mathbf{x})n(\mathbf{x}),
\end{equation}
where $n(\mathbf{x})=n(r,\theta,\phi)$ is the normalized density
distribution
\begin{equation}
\int n(\mathbf{x}) \der ^{3}\mathbf{x}=1.
\end{equation}
If assuming the angular distribution of samples is isotropic, the
radial distribution becomes $\tilde{n}(r)=4\pi r^{2} n(r)$. But
what we are interested is the ``true'' survey window function, by
taking into account of the effective sample depth
($R_{\rm{min}}$,$R_{\rm{max}}$) and partial sky coverage
($b_{\rm{cut}}$)\footnote{Since we are unclear about the real
angular selection function, we assume that it is uniformly
distributed above the sky-cut. If the angular distribution of the
survey is completely known, one can substitute it into
Eq.~(\ref{fourierwin1}) and calculate the corresponding filter.},
\begin{eqnarray}
\tilde{W}(k) & = & \frac{1}{\rm Vol}\int
W(\mathbf{x})n(\mathbf{x})
\cos(kr(\hat{\theta}\cdot \hat{\theta_{k}})) \nonumber \\
&\times&  r^{2}\sin(\theta)\der r \der \theta \der \phi,
\label{fourierwin1}
\end{eqnarray}
where we have used the real part of the plane wave $\exp({\rm i}
\mathbf{k} \cdot \mathbf{x})$ as the Fourier transform kernel.
Here we express the $\mathbf{k}=(k,\theta_{k},\phi_{k})$, and
$\textrm{Vol}$ is the volume of the survey. Substituting the
survey geometry and the cosine angle of $\theta$ and $\theta_{k}$
(i.e. $\hat{\theta} \cdot
\hat{\theta_{k}}=\cos\theta\cos\theta_{k}+\sin\theta\sin\theta_{k}\cos\phi$),
the radial distribution Eq.~(\ref{fourierwin1}) becomes
\begin{eqnarray}
\tilde{W}(k) & = &  \frac{1}{4 \pi \textrm{Vol}}
\int^{R_{\rm{max}}}_{R_{\rm{min}}} \tilde{n}(r) \nonumber \\
& \times & \cos\left(kr
\left(\cos\theta\cos\theta_{k}+\sin\theta\sin\theta_{k}\cos\phi
\right) \right) \nonumber \\  & \times &
\Theta(|\cos\theta|-\sin(b_{\rm{cut}})) \sin(\theta)\der r \der
\theta \der \phi,  \label{fourierwin2}
\end{eqnarray}
where $\Theta(x)$ is the Heaviside step function. The
$\textrm{Vol}$ in the denominator is the average factor, which
ensure the $\tilde{W}(k)$ is properly normalized
\begin{eqnarray}
\textrm{Vol} &= & \int W(\mathbf{x}) n(\mathbf{x}) \der ^{3}x
\nonumber
\\&=& \left[ \int^{R_{\rm{max}}}_{R_{\rm{min}}}  \tilde{n}(r) \der r
\right] \nonumber \\ &\times & \left[ \frac{1}{2} \int^{1}_{-1}
\Theta(|\cos\theta|-\sin(b_{\rm{cut}})) \right] \der  \cos\theta \nonumber \\
 &= & \left[ \int^{R_{\rm{max}}}_{R_{\rm{min}}}
\tilde{n}(r) \der r \right]\times (1-\sin(b_{\rm{cut}})).
\label{volfac}
\end{eqnarray}
Then we can substitute Eqs.~(\ref{fourierwin2}) and (\ref{volfac})
into Eq.~(\ref{sigmav3}) to calculate the rms of the bulk flow
velocity $V_{\rm{rms}}$ corresponding for the true survey volume
and samples.

\end{document}